\documentclass[graybox]{svmult}
\usepackage{mathrsfs}
\usepackage{amsmath}
\usepackage{mathptmx}       
\usepackage{helvet}         
\usepackage{courier}        
\usepackage{type1cm}        
\usepackage{makeidx}         
\usepackage{graphicx}        
\usepackage{multicol}        
\usepackage[bottom]{footmisc}
\usepackage{xcolor}          

\usepackage{bm}              

\usepackage{amssymb}

\DeclareFontFamily{OT1}{pzc}{}
\DeclareFontShape{OT1}{pzc}{m}{it}{<-> s * [1.200] pzcmi7t}{}
\DeclareMathAlphabet{\mathpzc}{OT1}{pzc}{m}{it}
\newcommand{\BEQ}{\begin{equation}}     
\newcommand{\BEA}{\begin{eqnarray}}
\newcommand{\BD}{\begin{displaymath}}
\newcommand{\EEQ}{\end{equation}}       
\newcommand{\EEA}{\end{eqnarray}}
\newcommand{\ED}{\end{displaymath}}
\newcommand{\vep}{\varepsilon}          
\renewcommand{\D}{{\rm d}}              
\newcommand{\II}{{\rm i}}               
\newcommand{\demi}{\frac{1}{2}}         
\newcommand{\wit}[1]{\widetilde{#1}}    
\newcommand{\wht}[1]{\widehat{#1}}      
\newcommand{\LLll}{\left\langle\!\left\langle}   
\newcommand{\RRrr}{\right\rangle\!\right\rangle} 

\renewcommand{\vec}[1]{\bm{#1}}          

\definecolor{gruen}{rgb}{0,0.625,0}     
\definecolor{rot}{rgb}{0.75,0,0}        
\definecolor{blau}{rgb}{0,0,0.75}       
\newcommand{\ROT}[1]{\textcolor{rot}{{\rm #1}}}	    
\newcommand{\VERT}[1]{\textcolor{gruen}{{\rm #1}}}	

\begin{document}

\title*{Schr\"odinger-invariance in non-equilibrium critical dynamics}
\titlerunning{Non-equilibrium critical dynamics}
\author{Malte Henkel and Stoimen Stoimenov}
\institute{ 
Malte Henkel$^{a,b}$ \at $^a$Laboratoire de Physique et Chimie Th\'eoriques {\small (CNRS UMR 7019)},
Universit\'e de Lorraine Nancy, B.P. 70239, F -- 54506 Vand{\oe}uvre-l\`es-Nancy Cedex, France\\
$^b$Centro de F\'{\i}sica T\'eorica e Computacional, Universidade de Lisboa,
P -- 1749-016 Lisboa, Portugal
\and Stoimen Stoimenov$^c$ \at $^c$Institute of Nuclear Research and Nuclear Energy, Bulgarian Academy of Sciences,
72 Tsarigradsko chaussee, Blvd., BG -- 1784 Sofia, Bulgaria
}
\maketitle

\abstract{The scaling functions of single-time and two-time correlators in systems undergoing non-equilibrium critical dynamics with dynamical exponent
$\mathpzc{z}=2$ are predicted from a new time-dependent non-equilibrium representation of the Schr\"odinger algebra. 
These explicit predictions are tested and confirmed in the ageing of several exactly solvable models. 
}

\section{Introduction: dynamical Schr\"odinger-symmetry} \label{sec:intro}

Many-body systems with strongly interacting degrees of freedom may display collective properties which need not be obvious from the
consideration of a single degree of freedom. While there only exist very few analytically solvable systems, they are often studied through
numerical simulations, which may require huge computational resources. Here we wish to explore the application of dynamical symmetries, not least motivated 
by the long-standing dream {\it ``we are looking forward to the day when Lie groups can be pushed to give \ldots the dynamics \ldots of a physical process.''} \cite{Gilm74}. 
Here we present a class of models whose dynamics can be described as following from the covariance of certain response functions under a new type of
non-equilibrium representation of the Schr\"odinger Lie algebra. 
In this paper, we shall consider systems quenched onto their critical point $T=T_c$; a companion paper 
will study dynamical symmetries in phase-ordering kinetics after a quench into the ordered phase with $T<T_c$ \cite{Stoi25V}. 

We shall begin with a short review of those field-theoretical ingredients 
we shall need to formulate our Schr\"odinger-covariance predictions (\ref{fonctions}) below. 
We shall work throughout in the continuum limit and
consider a coarse-grained `order-parameter' scaling operator $\phi(t,\vec{r})$ 
depending on the time $t$ and the space coordinates 
$\vec{r}$.\footnote{This follows the terminology from \cite{Card96}.} 
Averages of observables $\mathscr{A}=\mathscr{A}[\phi]$ 
should be found from a functional integral \cite{Jans76,Jans92,Domi76,Cala05,Taeu14}
\begin{subequations} 
\begin{align} \label{eq:1a}
\bigl\langle \mathscr{A}\bigr\rangle = \int \mathscr{D}\phi\mathscr{D}\wit{\phi}\; \mathscr{A}[\phi]\, e^{-{\cal J}[\phi,\wit{\phi}]}
\end{align}
\end{subequations}
where for systems far from equilibrium one has the Janssen-de Dominicis action 
\begin{align} \label{eq:2} 
{\cal J}[\phi,\wit{\phi}] &= \int \!\D t\D\vec{r}\: \left( \wit{\phi} \left( \partial_t - \Delta_{\vec{r}} - V'[\phi]\right)\phi - T \wit{\phi}^2 \right) 
\end{align}
with the interaction $V'[\phi]$ and the spatial laplacian $\Delta_{\vec{r}}$ (with usual re-scalings). 
The system is assumed to be in contact with a white-noise thermal bath of temperature $T$. 
The corresponding gaussian fluctuations can be integrated out which leads to the introduction of the response scaling operator $\wit{\phi}$ in (\ref{eq:1a}). 
Excluding any other source of noise implies that the system is brought onto a critical point, with temperature $T_c$, 
of the corresponding equilibrium system which is what we shall need later.

{\bf 1.} Define the {\em deterministic action} ${\cal J}_0[\phi,\wit{\phi}]  = \lim_{T\to 0} {\cal J}[\phi,\wit{\phi}]$ 
and {\em deterministic averages} $\bigl\langle \cdot \bigr\rangle_0$ as
\addtocounter{equation}{-2} 
\begin{subequations}
\addtocounter{equation}{1}  
\BEQ \label{eq:1b} 
\bigl\langle \mathscr{A}\bigr\rangle_0 = \int \mathscr{D}\phi\mathscr{D}\wit{\phi}\; \mathscr{A}[\phi]\, e^{-{\cal J}_0[\phi,\wit{\phi}]}
\EEQ
\end{subequations}
\addtocounter{equation}{1} 
Causality considerations \cite{Jans92,Cala05,Taeu14} or dynamical symmetries \cite{Barg54,Pico04} imply the Barg\-man su\-per\-se\-lec\-tion rules
\BEQ \label{Bargman}
\left\langle \overbrace{~\phi \cdots \phi~}^{\mbox{\rm ~~$n$ times~~}} 
             \overbrace{ ~\wit{\phi} \cdots \wit{\phi}~}^{\mbox{\rm ~~$m$ times~~}}\right\rangle_0 \sim \delta_{n,m}
\EEQ
for the deterministic averages. Only observables built from an equal number of order-parameters $\phi$ 
and conjugate response operators $\wit{\phi}$ can have non-vanishing deterministic averages. 
We shall find response functions $\langle \phi\wit{\phi}\rangle=\langle \phi\wit{\phi}\rangle_0$ and particularly correlators \cite{Pico04,Henk10}
\BEQ \label{corrFT}
C(t,s;r) 
= \bigl\langle \phi(t,\vec{r}+\vec{r}_0)\phi(s,\vec{r}_0) \bigr\rangle 
= T \int_0^{\infty} \!\D u \int_{\mathbb{R}^d} \!\D\vec{R}\: 
\left\langle \phi(t,\vec{r}+\vec{r}_0) \phi(s,\vec{r}_0) \wit{\phi^2}(u,\vec{R}) \right\rangle_0
\EEQ
by calculating $\langle\phi\phi\wit{\phi^2}\rangle_0$. Herein $\wit{\phi}_2 := \wit{\phi^2}$ is a composite scaling operator.

{\bf 2.} As already known to Jacobi and to Lie \cite{Duva24}, the Schr\"odinger group sends any solution of the diffusion equation 
$\bigl(2{\cal M}\partial_t - \partial_r^2\bigr)\phi=0$  onto another solution. 
For free fields or the $(1+1)D$ Calogero model, the deterministic action ${\cal J}_0$ has the Schr\"odinger group as a dynamical 
symmetry \cite{Henk03a,Shim21}. The Lie algebra $\mathfrak{sch}(1)$ is spanned by $\langle X_{\pm 1,0}, Y_{\pm\frac{1}{2}}, M_0\rangle$ 
(for simplicity in a $1D$ notation) where
\BEA
X_n &=& - t^{n+1}\partial_t - \frac{n+1}{2} t^n r\partial_r - (n+1) \delta t^n -\frac{n(n+1)}{4} {\cal M} t^{n-1} r^2 \nonumber \\
Y_m &=& - t^{m+\frac{1}{2}} \partial_r - \left( m + \frac{1}{2}\right) {\cal M} t^{m-\frac{1}{2}} \label{gl:5} \\
M_n &=& - t^n {\cal M} \nonumber
\EEA
act as super-operators on $\phi,\wit{\phi}$. An equilibrium scaling operator $\phi$ has a dimension $\delta$ and a mass ${\cal M}>0$. 
Covariance under the generators (\ref{gl:5}) implies \cite{Henk94} 
\BEA
\left\langle \phi_a(t_a,\vec{r}) \wit{\phi}_b(t_b,\vec{0})\right\rangle_0 
= \delta({\cal M}_a - {\cal M}_b)\, \delta_{\delta_a,\wit{\delta}_b}\: \Theta(t_{ab})\, t_{ab}^{-2\delta_a}\, 
\exp\left[ - \frac{{\cal M}_a}{2} \frac{\vec{r}^2}{t_{ab}} \right]
\label{gl:2points}
\EEA
\BEA
\lefteqn{\left\langle \phi_a(t_a,\vec{r}_a) \phi_b(t_b,\vec{r}_b)\wit{\phi}_c(t_c,\vec{r}_c)\right\rangle_0 = 
\delta({\cal M}_a + {\cal M}_b - {\cal M}_c)\, \Theta(t_{ac}) \Theta(t_{bc})\: }
\nonumber \\
&\times& t_{ac}^{-\delta_{ac,b}} t_{bc}^{-\delta_{bc,a}} t_{ab}^{-\delta_{ab,c}} \: 
\exp\left[ -\frac{{\cal M}_a}{2} \frac{\vec{r}_{ac}^2}{t_{ac}} - \frac{{\cal M}_b}{2} \frac{\vec{r}_{bc}^2}{t_{bc}}\right] 
\Phi_{ab,c}\left( \frac{\bigl[ \vec{r}_{ac}^2 t_{bc} - \vec{r}_{bc}^2 t_{ac} \bigr]^2}{t_{ab} t_{ac} t_{bc} }\right) ~~~
\label{gl:3points}
\EEA
Response operators have negative masses $\wit{\cal M}=-{\cal M}<0$ and $\wit{\delta}=\delta$. We abbreviate  
\BEQ
t_{ij} = t_i - t_j \;\; , \;\; \vec{r}_{ij} = \vec{r}_i - \vec{r}_j \;\; , \;\; \delta_{ij,k} = \delta_i + \delta_j - \delta_k
\EEQ
Causality in (\ref{gl:2points},\ref{gl:3points}) is guaranteed by the Heaviside functions $\Theta$ \cite{Henk03a}. 
The function $\Phi_{ab,c}$ is left undetermined.

{\bf 3.} The Schr\"odinger algebra contains time-translations $X_{-1}=-\partial_t$. Out-of-equi\-li\-brium, time-translation-invariance is no longer valid. 
We therefore use instead of the equilibrium representation $X^{\rm equi}$ (\ref{gl:5}) a non-equilibrium representation \cite{Henk25,Henk25c} 
\BEQ
X^{\rm equi} \mapsto X = e^{W(t)}\, X^{\rm equi}\, e^{-W(t)} \;\; , \;\; W(t) = \xi \ln t
\EEQ
For example, the generalised time-translation generator becomes $X_{-1}=-\partial_t +\frac{\xi}{t}$. A non-equilibrium scaling operator $\phi$ is
characterised by a pair $(\delta,\xi)$, besides its mass. The dilatation generator $X_0$ keeps its form and only modifies its scaling dimension
$\delta \mapsto \delta_{\rm eff} = \delta-\xi$. In this new representation the scaling operators become $\Phi(t) = t^{\xi} \phi(t) = e^{\xi \ln t}\phi(t)$.
The equilibrium response functions (\ref{gl:2points},\ref{gl:3points}) are mapped to non-equilibrium ones via 
(spatial arguments are suppressed for clarity)
\BEA
\bigl\langle \phi_a(t_a)\wit{\phi}_b(t_b)\bigr\rangle_0            
&~\mapsto~ & t_a^{\xi_a} t_b^{\wit{\xi}_b} \: \bigl\langle \phi_a(t_a)\wit{\phi}_b(t_b)\bigr\rangle_0 
\nonumber \\
\bigl\langle \phi_a(t_a)\phi_b(t_b)\wit{\phi}_c(t_c)\bigr\rangle_0 
&~\mapsto~ & t_a^{\xi_a} t_b^{\xi_b} t_c^{\wit{\xi}_c} \: \bigl\langle \phi_a(t_a) \phi_b(t_b)\wit{\phi}_c(t_c)\bigr\rangle_0 
\label{reponseHE}
\EEA

{\bf 4.} Furthermore, it can be shown that in the non-equilibrium representation, the non-linearity in the effective equation of motion
$\bigl(2{\cal M}\partial_t -\partial_{\vec{r}}^2\bigr)\phi=g \phi^3$ becomes irrelevant for large times provided $2\xi>1$ \cite{Stoi05,Henk25c}. 
Since the models we shall consider later all have linear equations of motion,  
this aspect is not going to be important for us and will be presented in more detail in \cite{Stoi25V}. 

Now combine the informations (\ref{corrFT},\ref{gl:2points},\ref{gl:3points},\ref{reponseHE}). We define the exponents 
\BEQ \label{2reponse-exp}
\frac{\lambda_C}{2} = \frac{\lambda_R}{2} = 2\delta -\xi \;\; , \;\; 1+a' = 2\delta \;\; , \;\; a'-a = \xi +\wit{\xi}
\;\; , \;\; b = \lambda_C -2\wit{\xi}_2  - \frac{d+2}{2}
\EEQ
Up to normalisations, the two-time response function is \cite{Henk06} 
\begin{subequations} \label{fonctions}
\begin{align} \label{2reponseR}
R(t,s;r) = \left\langle {\phi}(t,\vec{r}) {{\wit{\phi}}}(s,\vec{0})\right\rangle 
= s^{-1-a} \left( \frac{t}{s}\right)^{1+a'-\lambda_R/2} 
\left(\frac{t}{s} -1 \right)^{-1-a'} \exp\left[ -\frac{\cal M}{2} \frac{r^2}{t-s} \right] 
\end{align}
whereas the single-time correlator takes the form \cite{Henk25d}
\begin{align} \label{1correlateurC}
C(t;\vec{r}) = \bigl\langle \phi(t,\vec{r}) \phi(t,\vec{0}) \bigr\rangle
= t^{-b} e^{-\frac{\cal M}{4}\frac{\vec{r}^2}{t}}\, U\left( 2\wit{\xi}_2+1, 4\wit{\delta}_2-\frac{d}{2};\frac{\cal M}{4}\frac{\vec{r}^2}{t}\right)
\end{align}
where $U$ is a Kummer/Tricomi confluent hypergeometric function \cite{Abra65}. There is the constraint $\wit{\delta}_2=\wit{\delta}=\delta$.
Finally, the two-time auto-correlator is \cite{Henk25d} 
\begin{align} \label{2correlateurC}
& C(ys,s) = \bigl\langle \phi(ys,\vec{0}) \phi(s,\vec{0}) \bigr\rangle 
       \:=\: s^{-b}\, y^{\xi+d/2-2\wit{\delta}_2} \big(y+1\bigr)^{-d/2-\nu} \bigl(y-1\bigr)^{\nu} \nonumber \\
&  ~~\times F_1\left( 2\wit{\xi}_2+1,2\wit{\delta}_2-d/2,d/2+\nu;2+d/2+2\wit{\xi}_2-2\wit{\delta}_2;\frac{1}{y},\frac{2}{y+1}\right)
\end{align}
where $F_1$ is an Appell function \cite{Prud3}. This depends on the additional hypothesis that 
$\Psi(Y) := \int_{\mathbb{R}^d} \!\D\vec{P}\: e^{-\demi{\cal M}\vec{P}^2}\Phi_{\phi\phi,\wit{\phi}_2}\bigl(\vec{P}^2 Y\bigr) \stackrel{!}{=} \Psi_{\infty} Y^{\nu}$ 
becomes a simple power where $\Phi_{\phi\phi,\wit{\phi}_2}$ was left undetermined in (\ref{gl:3points}). 
It will turn out to hold true in all models we shall consider below. 
In the special case where $2\wit{\delta}_2=\frac{d}{2}$, the second argument of $F_1$ 
vanishes so that we can use $F_1\bigl(a,0,b;c;w,z) = {}_2F_1\bigl(a,b;c;z)$ \cite{Prud3}. Then
\begin{align} \label{2correlateurC-simp}
C(ys,s) &= \frac{s^{-b} y^{\xi} \bigl(y-1\bigr)^{\nu}}{\bigl(y+1\bigr)^{d/2+\nu}} \, 
{}_2F_1\left( 2\wit{\xi}_2+1, \frac{d}{2}+\nu;2\wit{\xi}_2+2;\frac{2}{y+1}\right) \;\; ; \;\; \wit{\delta}_2=\frac{d}{4} 
\end{align}
\end{subequations}
The functional forms of the correlators depend only on properties of the composite response field $\wit{\phi}_2 =\wit{\phi^2}$, see \cite{Anti25}. 
Eqs.~(\ref{fonctions}) realise the long-standing ambition \cite{Gilm74} to predict the form of time-dependent observables 
in interacting many-body systems, out of equilibrium,  merely from a dynamical symmetry. 
Information on a specific model only enters through the values of the parameters $(\xi,\wit{\xi},\wit{\xi}_2,\wit{\delta}_2)$ and possibly $\nu$. 

\section{Physical ageing: non-equilibrium critical dynamics} \label{sec:age}

The theory of section~\ref{sec:intro} will be applied to ageing systems \cite{Stru78,Puri09,Henk10,Taeu14,Cugl15,Vinc24}.  
In classical dynamics, {\em physical ageing} arises in a many-body system which is prepared in an initially fully disordered state 
(hence $\langle \phi(t,\vec{r})\rangle=\langle\phi(0,\vec{r})\rangle=0$) 
and is then instantaneously quenched onto its equilibrium critical point
$T_c$. The critical fluctuations in the system lead to the formation of correlated clusters \cite{Alme25} with a time-dependent length scale $\ell(t)$. 
If for late times $\ell(t)\sim t^{1/\mathpzc{z}}$, this defines the dynamical exponent $\mathpzc{z}$. 
Here, we shall only consider systems where $\mathpzc{z}=2$ which implies that any conservation laws on the order-parameter $\phi$ are excluded. 
The relaxation-time towards equilibrium becomes formally infinite, and one speaks of {\em non-equilibrium critical dynamics} \cite{Godr02}.
For quenches to $T<T_c$ one obtains {\em phase-ordering kinetics} treated in \cite{Stoi25V}.
Physical observables include the two-time correlator $C$ and the two-time response $R$ with respect to an external perturbing field $h$, defined as 
\begin{subequations} \label{gl:13}
\begin{align} 
C(t,s;{r}) &:= \left\langle \phi(t,\vec{r}) \phi(s,\vec{0}) \right\rangle = s^{-b} F_C\left(\frac{t}{s};\frac{r}{s^{1/2}}\right) \label{gl:13C}  \\ 
R(t,s;{r}) &:= \left.\frac{\delta\bigl\langle \phi(t,\vec{r})\bigr\rangle}{\delta h(s,\vec{0})}\right|_{h=0}
                 =\left\langle \phi(t,\vec{r}) \wit{\phi}(s,\vec{0}) \right\rangle
                 = s^{-1-a} F_R\left(\frac{t}{s},\frac{r}{s^{1/2}}\right) \label{gl:13R} 
\end{align} 
and we reused the response scaling operator $\wit{\phi}$ \cite{Jans76,Domi76}. 
We always assume spatial translation- and rotation-invariance such that $\vec{r}\mapsto r = |\vec{r}|$. 
The associated scaling form are valid for $t,s\gg \tau_{\rm micro}$ and $t-s\gg\tau_{\rm micro}$ and are illustrated in figures~\ref{fig1} and~\ref{fig2} 
(specifically, for the voter model, to be defined below).

\begin{figure}[tb]
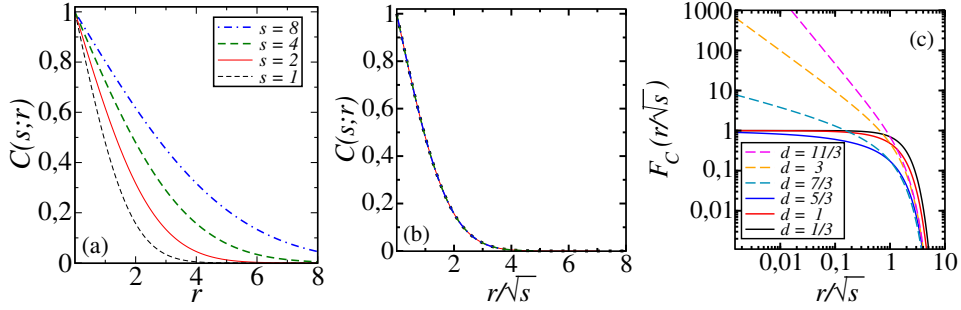

\begin{center}
\includegraphics[width=.36\hsize]{actes_varnaLT16-voter_vieilC1_bis.eps}~
\includegraphics[width=.34\hsize]{actes_varnaLT16-voter_vieilC1-skal_bis.eps}~
\includegraphics[width=.36\hsize]{actes_varnaLT16-voter_vieilC1-ska_dim_bis.eps}~
\end{center}
\caption[fig1]{\small Ageing of the single-time correlator $C(s;r)$ in the voter model.
(a) Dependence of $C(s;r)$ on $r$ for several values of the time $s$ for $d=1$ and (b) the data collapse when replotted over against 
$r/\sqrt{s\,}$. (c) Form of the scaling function $F_C(1,r/\sqrt{s\,})=s^b C(s;r)$ for
$d=[\frac{1}{3},1,\frac{5}{3}]$ (full lines) from top to bottom and for
$d=[\frac{7}{3},3,\frac{11}{3}]$ (dashed lines) from bottom to top.\label{fig1}}
\end{figure}

In figure~\ref{fig1} a typical single-time correlator $C(s;\vec{r})=C(s,s;\vec{r})$ is shown. 
First, figure~\ref{fig1}a shows that ageing is a slow dynamics, as with increasing age $s$, the correlator $C(s;\vec{r})$ decays more slowly. 
Second, there is no time-translation-invariance since for each $s$ the is a distinct curve $C(s;\vec{r})$.
Third, there is dynamical scaling since the same data, replotted over against $r/\sqrt{s}$, collapse onto a single curve, see figure~\ref{fig1}b. 
Several examples of this
scaling function are shown in figure~\ref{fig1}c for various values of the dimension $d$. 
The same observations are made in figure~\ref{fig2} for the two-time auto-correlator $C(t,s) := C(t,s;0)$. 
First, the dynamics is slow, since with increasing waiting time $s$, the auto-correlator $C(s+\tau,s)$ decays more slowly, see figure~\ref{fig2}a. 
Second, there is no time-translation-invariance, since for each value of $s$, there is a distinct curve. 
Third, there is dynamical scaling, proven by the data collapse when the same data are replotted in figure~\ref{fig2}b over against $y=t/s$. 
This demonstrates the three defining properties of ageing \cite{Stru78,Henk10} for the correlators and can be done similarly for the response as well. This
is expressed formally in the scaling form (\ref{gl:13}). The auto-correlator and -response scaling functions are asymptotically algebraic
\begin{align}
f_C(y) = F_C(y;0) \sim y^{-\lambda_C/2} \;\; , \;\; f_R(y) = F_R(y;0) \sim y^{-\lambda_R/2} \;\; ; \;\; y\gg 1
\end{align}
\end{subequations}
which define the {\em auto-correlation exponent} $\lambda_C$ (see figure~\ref{fig2}c) and the {\em auto-response exponent} $\lambda_R$ \cite{Huse89}. 
For spatially short-ranged initial correlations, one expects $\lambda=\lambda_C=\lambda_R$. 
In non-equilibrium critical dynamics, the {\em ageing exponent}s $b=a=(d-2+\eta)/\mathpzc{z}$ where $\eta$ is an equilibrium critical exponent. 

\begin{figure}[tb]
\begin{center}
\includegraphics[width=.355\hsize]{actes_varnaLT16-voter_vieilC2_bis.eps}~
\includegraphics[width=.34\hsize]{actes_varnaLT16-voter_vieilC2-ska_bis.eps}~
\includegraphics[width=.34\hsize]{actes_varnaLT16-voter_vieilC2-ska_dim_bis.eps}~
\end{center}
\caption[fig2]{\small Ageing of the two-time auto-correlator $C(t,s)$ in the voter model.
(a) Dependence of $C(s+\tau,s)$ on $\tau$ for several waiting times $s$ for $d=1$ and (b) the data collapse when replotted over against 
$y=t/s$. (c) Form of the scaling function $f_C(y)=s^b C(ys,s)$ for
$d=[\frac{1}{3},1,\frac{5}{3}]$ (full lines) from top to bottom and for
$d=[\frac{7}{3},3,\frac{11}{3}]$ (dashed lines) from bottom to top. \label{fig2}}
\end{figure}

Figures~\ref{fig1}c and~\ref{fig2}c show that the form of these scaling functions is distinct for $d<d^*$ and for $d>d^*$. 
They are expected to be universal, that is independent of `specific' details such as lattice structure or details of the interactions. 
Field-theory methods such as the $\vep$-expansion permit to compute their forms \cite{Cala05,Taeu14}, but we shall strive to find those by comparing with the
results (\ref{fonctions}) of Schr\"odinger symmetry reviewed in section~\ref{sec:intro}.  

\section{Models} \label{sec:models}

\subsection{Voter model}

The {voter model} is a paradigm for non-equilibrium ageing \cite{Dorn01}. On a $d$-dimensional hyper-cubic lattice $\Lambda\subset\mathbb{Z}^d$, it is defined
in terms of spin variables $\sigma_{\vec{n}}=\pm 1$ attached to each site $\vec{n}\in\Lambda$. Each configuration $\{\sigma\}$ of spins occurs with probability
$P(\{\sigma\},t)$ and its evolution is described by a master equation 
\BEQ
\partial_t P\bigl( \{\sigma\},t) = \sum_{\{\sigma'\}} \left[ w_{\{\sigma' \}\to \{\sigma \}} P\bigl(\{\sigma'\},t) 
-w_{\{\sigma \}\to \{\sigma' \}} P\bigl(\{\sigma\},t) \right]
\EEQ
In the {\em voter model}, transitions between configurations occur via single spin flips at site $\vec{n}$ and with the rates \cite{Ligg85,Tome01,Krap10}
\BEQ \label{gl:voter-rates}
w_{\{\sigma \}\to \{\sigma' \}} ~~\mapsto~~  
w_{\vec{n}}\bigl(\{\sigma \}\bigr) =  \demi\left( 1 - \frac{1}{2d} \sigma_{\vec{n}} \sum_{\vec{m}(\vec{n})} \sigma_{\vec{m}} \right)
\EEQ
where $\vec{m}(\vec{n})$ are the nearest-neighbour sites with respect to the site $\vec{n}\in\Lambda$. 
For $d=1$ this reduces to the Glauber-Ising chain \cite{Glau63} at $T=0$ 
and only in this special case do the rates $w_{\vec{n}}$ obey the detailed balance condition 
$w_{\{\sigma' \}\to \{\sigma \}} P_{\rm eq}\bigl(\{\sigma'\}) =w_{\{\sigma \}\to \{\sigma' \}} P_{\rm eq}\bigl(\{\sigma\})$ 
which guarantees the relaxation towards an equilibrium state with a Boltzmann distribution $P_{\rm eq}(\{\sigma\})$. 
The rates (\ref{gl:voter-rates}) lead to linear and closed equations of motion for correlators and responses. 
Hence the voter model is exactly solvable in all dimensions $d>0$ \cite{Frac97} and we find (after rescaling) \cite{Henk25f}
\begin{subequations} \label{gl:voter-resultat}
\begin{align}
R(ys,s;\vec{r}) &= \left\{ \begin{array}{ll}
\frac{\mathfrak{R}_0}{s}       \bigl(y-1\bigr)^{-d/2} \exp\left[ -\demi \frac{r^2}{s(y-1)}\right] & \mbox{\rm ~~;~ if $d<2$} \\
\frac{\mathfrak{R}_0}{s\ln s}  \bigl(y-1\bigr)^{-1}   \exp\left[ -\demi \frac{r^2}{s(y-1)}\right] & \mbox{\rm ~~;~ if $d=2$} \\
\frac{\mathfrak{R}_0}{s^{d/2}} \bigl(y-1\bigr)^{-d/2} \exp\left[ -\demi \frac{r^2}{s(y-1)}\right] & \mbox{\rm ~~;~ if $d>2$} 
\end{array} \right.
\end{align}
with a $d$-dependent normalisation $\mathfrak{R}_0$, and 
\begin{align} \label{gl:voter-resultatC1}
C(s;r) &= \left\{ \begin{array}{ll}
\frac{1}{\Gamma(1-\frac{d}{2})}\,e^{-r^2/4t}\, U\left(\frac{d}{2},\frac{d}{2};\frac{r^2}{4t}\right) & \mbox{\rm ~~;~ if $d<2$} \\
\frac{1}{\ln s}\, e^{-r^2/4t}\, U\left(1,1;\frac{r^2}{4t}\right) & \mbox{\rm ~~;~ if $d=2$} \\
\frac{\mathfrak{C}_0}{2^{d-2}\Gamma(\frac{d}{2}-1)}\, s^{1-d/2}\,e^{-r^2/4t}\, U\left(1,\frac{d}{2};\frac{r^2}{4t}\right) & \mbox{\rm ~~;~ if $d>2$} \\
\end{array} \right.
\end{align}
where $U$ is a Kummer/Tricomi hypergeometric function \cite{Abra65} and $\mathfrak{C}_0$ a normalisation constant and finally 
\begin{align} \label{gl:voter-resultatC2}
C(ys,s) &= \left\{ \begin{array}{ll}
\frac{1}{\Gamma(1-\frac{d}{2})\Gamma(1+\frac{d}{2})} \left(\frac{2}{y+1}\right)^{d/2}{}_2F_1\left(\frac{d}{2},\frac{d}{2};\frac{d}{2}+1;\frac{2}{y+1}\right) & \mbox{\rm ~~;~ if $d<2$} \\
\frac{\mathfrak{C}_0}{\ln s} \left(\frac{2}{y+1}\right){}_2F_1\left(1,1;2;\frac{2}{y+1}\right)                                                               & \mbox{\rm ~~;~ if $d=2$} \\
\frac{\mathfrak{C}_0}{2^{d-1}\Gamma(\frac{d}{2})}s^{1-d/2}\left(\frac{2}{y+1}\right)^{d/2} {}_2F_1\left(\frac{d}{2},1;2;\frac{2}{y+1}\right)                 & \mbox{\rm ~~;~ if $d>2$} \\
\end{array} \right.
\end{align}
\end{subequations}
All known results for $d=1,2,3$, see \cite{Droz89,Krap92,Bray97,Frac97,Godr00a,Lipp00,Sast03,Corb24a,Corb24b,Corb24c,Corb24e,Henk25b}, are contained herein as special cases. 
The analytic results (\ref{gl:voter-resultat}) prove the scaling expectations (\ref{gl:13}) of ageing and from (\ref{gl:voter-resultatC1},\ref{gl:voter-resultatC2}) we have
the analytic proof of all observations made on the scaling functions in figures~\ref{fig1}c and~\ref{fig2}c. The upper critical dimension is $d^*=2$. 

Finally, the comparison of (\ref{gl:voter-resultat}) with non-equilibrium Schr\"odinger-invariance 
(\ref{fonctions}) gives a complete and consistent agreement between all three observables 
so that we can identify the universal scaling dimensions $(\xi,\wit{\xi},\wit{\xi}_2,\wit{\delta}_2)$ along with $\nu=0$ and the non-universal ${\cal M}=1$. 
The results are listed in table~\ref{tab:1}, where we included the Glauber-Ising chain \cite{Bray97}, quenched to $T=0$, as a separate entry as well. 

\subsection{Spherical model} 

The critical {\em spherical model} is a classic model of interacting spins which can also be obtained as the $N\to\infty$ limit of the O$(N)$ Heisenberg model. 
Heuristically, it may be introduced by relaxing the constraint 
$\sigma_{\vec{n}}=\pm 1$ of the Ising model spin variables, with the purpose of obtaining an exactly solvable model. 
Spherical spins are continuous $S_{\vec{n}}\in\mathbb{R}$ for all $n\in\Lambda$ and obey the mean spherical constraint 
$\sum_{\vec{n}\in\Lambda} \langle S_{\vec{n}}^2\rangle =|\Lambda|$, where $|\Lambda|$ is the number of
sites of the lattice $\Lambda\subset\mathbb{Z}^d$. At equilibrium, the nearest-neighbour model has a phase-transition for $d>2$, with 
the critical temperature $T_c^{-1}=\int_0^{\infty}\!\D u\: \bigl( e^{-2u}I_0(2u)\bigr)^d>0$ where 
$I_0$ is a modified Bessel function \cite{Abra65}. 
In the continuum limit, the local magnetisation $S(t, \vec{r})$ obeys the Langevin equation \cite{Godr00b} 
\BEQ
\partial_t S(t,\vec{r}) = \Delta_{\vec{r}} S(t,\vec{r}) - \mathfrak{z}(t) S(t,\vec{r}) + \eta(t,\vec{r}) \;\; , \;\;
\langle \eta(t,\vec{r})\eta(0,\vec{0})\rangle = 2T_c \delta(t)\delta(\vec{r})
\EEQ
with the spatial laplacian $\Delta_{\vec{r}}$ and a centred gaussian noise $\eta(t,\vec{r})$. 
The Lagrange multiplier $\mathfrak{z}(t)$ enforces the mean spherical constraint. 
To solve the model for any $d>0$, define
\BEQ 
g(t) = \exp\left( 2 \int_0^t \!\D\tau\: \mathfrak{z}(\tau) \right)
= \bigl(e^{-4t}I_0(4t)\bigr)^d + 2T \int_0^t \!\D\tau\: g(t-\tau) \bigl( e^{-4\tau}I_0(4\tau)\bigr)^d 
\EEQ
and this Volterra integral equation follows from the spherical constraint \cite{Godr00b}. Solving this leads to the long-time behaviour, for $t\gg 1$
\BEQ \label{gl:19}
g(t) \simeq -\frac{1}{2T_c} \delta(t) + g_{d}\, t^{\digamma} \;\; , \;\; 
\digamma = \left\{ \begin{array}{ll} d/2-2 & \mbox{\rm ~~;~ if $2<d<4$} \\
                                     0     & \mbox{\rm ~~;~ if $4<d$}
                   \end{array} \right.
\EEQ
which now permits to work out the observables. The singular term in this is needed in the calculation of correlators, 
while it plays no r\^ole for the long-time behaviour of response functions. 
One has, up to normalisation for the two-time response for $d\ne 4$ \cite{Godr00b} 
\begin{subequations}
\begin{align}
R(ys,s;r) = \frac{s^{-d/2}}{(4\pi)^{d/2}}\, \frac{y^{-\digamma/2}}{\bigl(y-1\bigr)^{d/2}} \exp\left[ -\frac{1}{4}\frac{r^2}{s(y-1)}\right]
\end{align}
The single-time correlator is, with (\ref{gl:19}) and recast in the form most useful to us 
\begin{align}
C(s;r) = \mathfrak{C}_0\: s^{1-d/2}\, e^{-\frac{r^2}{8t}}\, U\left( \digamma+1,\frac{d}{2};\frac{r^2}{8t}\right)
\end{align}
and the two-time auto-correlator is
\begin{align}
C(ys,s) = \frac{2T_c\:s^{1-d/2}}{(2\pi)^{d/2-1}(\digamma+1)}  \frac{y^{-\digamma/2}}{\bigl(y+1\bigr)^{d/2}}\, 
{}_2F_1\left(\frac{d}{2},\digamma+1;\digamma+2;\frac{2}{y+1}\right)
\end{align}
\end{subequations}
which we recast in the form most useful for our purposes.\footnote{We use the identity $\lim_{\vep\to 0}U(a-1+\vep,a;z)=z^{1-a}$ for $a\not\in\mathbb{N}$.}  
There is complete agreement with the non-equilibrium prediction (\ref{fonctions}). 
We identify the scaling dimensions listed in table~\ref{tab:1} and notice that $\nu=0$ and ${\cal M}=\demi$.

\subsection{Edwards-Wilkinson model} \label{subsec:ew} 

This model \cite{Edwa82,Bara95} describes the growth of a surface whose height takes the r\^ole of the order-parameter $\phi(t,\vec{r})$. 
It satisfies a noisy diffusion equation
$\bigl(\partial_t - \Delta_{\vec{r}}\bigr)\phi(t,\vec{r})=\eta(t,\vec{r})$ where the gaussian white noise has the second moment 
$\langle \eta(t,\vec{r})\eta(t',\vec{r}')\rangle=2\delta(t-t')\delta(\vec{r}-\vec{r}')$. 

Up to normalisation, for all $d>0$ the exact results \cite{Roet06} for the two-time observables can be recast, after re-scaling, as follows 
\begin{subequations} \label{gl:ew}
\begin{align}
R(ys,s;\vec{r}) = \mathfrak{R}_0\, s^{-d/2} \bigl( y-1\bigr)^{-d/2} \exp\left[ -\frac{1}{4} \frac{r^2}{s(y-1)}\right]
\end{align}
\begin{align}
C(ys,s) = \mathfrak{C}_0\, s^{1-d/2} \left(\frac{2}{y+1}\right)^{d/2} {}_2F_1\left(\frac{d}{2},1;2;\frac{2}{y+1}\right)
\end{align}
which includes the logarithmic forms for $d=2$ \cite{Roet06}. 
To this we add the single-time correlator \cite{Henk17c,Duva24}, recast as  
\begin{align}
C(s;r) = \mathfrak{C}_0\, s^{1-d/2} e^{-r^2/4s}\, U\left(1,\frac{d}{2};\frac{r^2}{8s}\right)
\end{align}
\end{subequations}
This agrees with the generic expectation (\ref{fonctions}) and the resulting scaling dimensions are listed in table~\ref{tab:1}. 
Indeed, it is known since a long time that the {\sc ew} model is described by the equilibrium representation (\ref{gl:5}) 
of the Schr\"odinger algebra with $\xi=\wit{\xi}=\wit{\xi}_2=0$, $\nu=0$ and ${\cal M}=\demi$ \cite{Roet06}. 

\subsection{Arcetri model} 

\begin{figure}[tb]
\sidecaption
\includegraphics[width=.5\hsize]{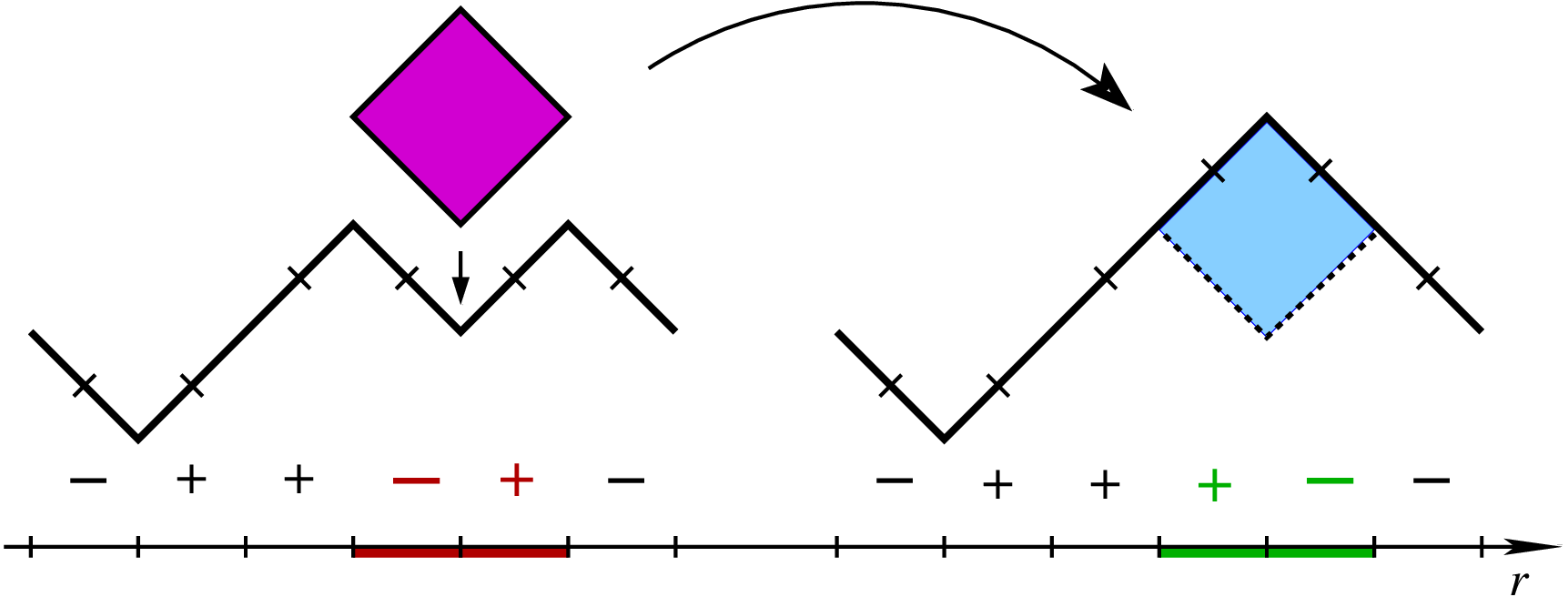}~
\caption[fig3]{\small Evolution step of a growing interface with the {\sc rsos} constraint. \label{fig3}}
\end{figure}

Before we continue with the discussion of another model of surface growth, we need more background \cite{Bara95,Halp95}. The surface will be described, in the
continuum limit, by a height profile $h(t,\vec{r})$ and this is usually assumed to be given in terms of a Kardar-Parisi-Zhang ({\sc kpz}) equation
\BEQ
\partial_t h = \Delta_{\vec{r}}h +\frac{\mu}{2} \bigl( \nabla_{\vec{r}} h\bigr)^2 +\eta
\EEQ
with a white noise $\eta$ and the constant $\mu$. For $\mu=0$, one is back to the {\sc ew}-model of subsection~\ref{subsec:ew}. On a lattice, one often
uses a representation where the heights on nearest neighbouring sites $\vec{j},\vec{j}'$ obey a {\sc rsos} constraint $h_{\vec{j}}-h_{\vec{j}'}=\pm 1$,
see figure~\ref{fig3} for a $1D$ illustration. Evolution of the surface is only possible via the specific step indicated. In terms of the slopes, this
means that only the totally asymmetric move $\ROT{\bf{-}{\bf+}}\to \VERT{{\bf+}{\bf-}}$ is admissible. 

We use that all nearest-neighbour slopes are restricted to the values $\pm 1$ by the {\sc rsos} condition. We follow \cite{Henk15,Dura17} and relax
this in a way quite analogous to the spherical model. First, for the heights we assume a Langevin equation
\BEQ \label{gl:Arcetrih}
\partial_t h(t,\vec{r}) = \Delta_{\vec{r}}h(t,\vec{r}) +\mathfrak{z}(t) h(t,\vec{r}) + \eta(t,\vec{r})
\EEQ
with the white noise $\eta$ and the Lagrange multiplier $\mathfrak{z}(t)$. We assume an initially uncorrelated surface
\BEQ \label{gl:Arcetrih_ini}
\LLll h(0,\vec{r}) \RRrr = H_0 \;\; , \;\;
\LLll h(0,\vec{r})h(0,\vec{r}')\RRrr - \LLll h(0,\vec{r})\RRrr \LLll h(0,\vec{r}')\RRrr = H_1 \delta(\vec{r}-\vec{r}')
\EEQ
It is useful to consider the slopes $u_a(t,\vec{r})=\partial_a h(t,\vec{r})$, their Langevin equation
\begin{subequations} \label{gl:Arcetriu}
\BEQ \label{gl:Arcetriu1}
\partial_t u_a(t,\vec{r}) = \Delta_{\vec{r}}u_a(t,r) + \mathfrak{z}(t) u_a(t,\vec{r}) + \partial_a \eta(t,\vec{r})
\EEQ
and which are required to obey the mean spherical constraint (${\cal B}=[\pi,\pi]^d$ is the Brillouin zone) 
\BEQ \label{gl:Arcetriu2}
\sum_{a=1}^{d} \frac{1}{(2\pi)^d} \int_{\cal B} \!\D \vec{r}\: \LLll\, \left\langle u_{a}(t,\vec{r})^2 \right\rangle\, \RRrr = d
\EEQ
\end{subequations}
This defines the {\em Arcetri model} \cite{Henk15,Dura17} which stands with respect to the {\sc kpz} 
equation in about the same kind of relationship as stands the spherical model
with respect to the Ising model.\footnote{More precisely, this model is referred to as {\sc Arcetri 1H} in \cite{Dura17}.} 
When concentrating on the slope variables $\vec{u}(t,\vec{r})$, 
it can be shown that the $1D$ Arcetri model is identical \cite{Henk15} to the $p=2$ spherical spin glass \cite{Cugl95}. 
Furthermore, the $d$-dimensional Arcetri model gives the same correlators and responses as the spherical model in $d+2$ dimensions \cite{Henk15}. 
The exact solution follows closely the one of the spherical model, but we give a few more steps since the model is less well-known. 
In Fourier space (\ref{gl:Arcetriu1}) turns into
\BEQ \label{gl:AuFourier}
\wht{u}_a(t,\vec{p}) = \wht{u}_a(0,\vec{p}) \frac{e^{-2\omega(\vec{p})t}}{\sqrt{g(t)\,}} 
                       +\II \int_0^{t}\!\D\tau\: \sqrt{\frac{g(\tau)}{g(t)}\,}\, p_a\, e^{-2\omega(\vec{p})(t-\tau)}
\EEQ
with $\omega(\vec{p})=\sum_{a=1}^d \bigl(1-\cos p_a\bigr)$ and we also define
\BEQ
d\, g(t) = d e^{-2\int_0^t \!\D t'\: \mathfrak{z}(t')} = H_1 f(t) + 2T \int_0^{t} \!\D\tau\: f(\tau) g(t-\tau)
\EEQ
where the Volterra integral equation follows from the constraint (\ref{gl:Arcetriu2}), by working out the correlation averages obtained from
(\ref{gl:AuFourier}), and where
\BEQ
f(t) := \frac{1}{(2\pi)^d} \int_{\cal B} \!\D\vec{p}\:  \vec{p}^2\, e^{-4\omega(\vec{p})t} 
= d\frac{e^{-4t} I_1(4t)}{4t} \left( e^{-4t} I_0(4t)\right)^{d-1} 
\EEQ
with the modified Bessel functions $I_{0,1}$ \cite{Abra65}. The critical point is
located at $T_c^{-1}=\demi\int_0^{\infty}\!\D u\: e^{-dt} u^{-1} I_1(u) I_0^{d-1}(u)>0$ for all $d>0$. For late times $t\gg 1$, and at $T=T_c$, we find \cite{Henk15} 
\BEQ \label{gl:Adigamma}
g(t) = -\frac{H_1}{2T_c} \delta(t) + g_d\, t^{\digamma} \;\; , \;\; \digamma = \left\{ \begin{array}{ll} d/2-1 & \mbox{\rm ~~;~ if $d<2$} \\
                                                                                                       0     & \mbox{\rm ~~;~ if $d>2$} 
                                                                                     \end{array} \right. 
\EEQ
where the singular term is important for the calculation of correlators and the known constant $g_d$ will drop out for almost all observables.

We can now list the non-equilibrium averages for the Arcetri model quenched onto its critical point $T_c$ \cite{Henk15}. 
We merely use the slightly modified definitions, with respect to the spin models discussed before, 
as they arise in terms of the height variables in the Arcetri model.
\begin{subequations}
For the response of the height $h(t,\vec{r})$ with respect to an external source, we have, with $\digamma$ taken from  (\ref{gl:Adigamma}) 
\begin{align}
R(ys,s;r) = \left.\frac{\delta \langle h(t,\vec{r})\rangle}{\delta j(s,\vec{0})}\right|_{j=0} 
= \frac{s^{-d/2}}{(4\pi)^{d/2}}\, \frac{y^{-\digamma/2}}{\bigl(y-1\bigr)^{d/2}} \exp\left[ -\frac{1}{2}\frac{r^2}{s(y-1)}\right]
\end{align}
The connected single-time correlator is, averaged over both the initial state and the thermal bath, where $\overline{h}(s)$ is the spatially averaged height, 
\begin{align}
C(s;r) &= \LLll \,\left\langle \bigl(h(s,\vec{r})-\overline{h}(t)\bigr)\bigl(h(s,\vec{0})-\overline{h}(s)\bigr) \right\rangle\,\RRrr \nonumber \\
&= \frac{2T_c\:s^{1-d/2}}{(8\pi)^{d/2}\Gamma(1+\digamma)}\: e^{-\frac{r^2}{2t}}\, U\left( \digamma+1,\frac{d}{2};\frac{r^2}{4s}\right)
\end{align}
where we recast the explicit results in the form most useful to us.\footnote{Herewith we correct an important copying error in \cite[eq. (A.7)]{Henk15}.} 
Finally, the connected two-time auto-correlator is after averaging 
\begin{align}
C(ys,s) &= \LLll \,\left\langle \bigl(h(ys,\vec{0})-\overline{h}(ys)\bigr)\bigl(h(s,\vec{0})-\overline{h}(s)\bigr) \right\rangle\,\RRrr \nonumber \\
& = \frac{2T_c\:s^{1-d/2}}{(4\pi)^{d/2}(\digamma+1)}  \frac{y^{-\digamma/2}}{\bigl(y+1\bigr)^{d/2}} \, 
{}_2F_1\left(\frac{d}{2},\digamma+1;\digamma+2;\frac{2}{y+1}\right)
\end{align}
\end{subequations}
This permits to read off the exponents which describe the non-equilibrium ageing in this model. 
Since these are well-known, our interest centres on the form of the various
scaling function and the comparison with the predictions (\ref{fonctions}) of non-equilibrium Schr\"odinger-invariance. 
Indeed, there is an overall agreement so that we can identify
the various scaling dimensions which are listed in table~\ref{tab:1}.  We also notice that $\nu=0$ and the non-universal ${\cal M}=1$. 
For $d>2$, the Arcetri model is in the same universality class as its mean-field approximation, the {\sc ew}-model. 

In the context of interface growth, we add a last comment. An important observable is the interface width
\BEQ \label{gl:Aw}
w^2(t) = C(t,t) = C(t;\vec{0}) \sim t^{2\beta} = \left\{ \begin{array}{ll} d/2-1 & \mbox{\rm ~~;~ if $d<2$} \\
                                                                                                       0     & \mbox{\rm ~~;~ if $d>2$} 
                                                                                     \end{array} \right. 
\EEQ
where $\beta$ is usually called the {\em growth exponent} \cite{Bara95,Halp95}. 
Therefore, we read off for the ageing exponent $b=-2\beta=\digamma$ (there is no obvious expectation for the exponent $a$). 
On the other hand, we can also work out the height
profile direct by returning to (\ref{gl:Arcetrih}) with $\mathfrak{z}(t)$ already known. 
With the initial condition (\ref{gl:Arcetrih_ini}) of an initially flat and uncorrelated 
interface, the straightforward integration leads to
\BEQ \label{gl:Ah}
\LLll \, \left\langle\: h(t,\vec{r})\:\right\rangle\,\RRrr = \frac{H_0}{\sqrt{g(t)\,}} \sim t^{-\digamma/2} \sim t^{\beta}
\EEQ
where we implicitly assume to be in a co-moving frame with the mean velocity of the interface. 
In the {\sc kpz} equation, this relation between growth of the interface width (\ref{gl:Aw}) 
and the anomalous height growth (\ref{gl:Ah}) is called the {\em {\sc kpz} ansatz} \cite{Bara95}.
It is thought to be brought about by the fluctuations in the {\sc kpz} equation. 
It is satisfying that the {\sc kpz} ansatz is satisfied in the Arcetri model for $d<2$ as well, 
where fluctuations lead to non-trivial values of the non-equilibrium exponents.

\subsection{Bosonic contact process}

Particle-reaction models are widely studied in various branches of non-equilibrium systems. 
Here we concentrate on the bosonic variant where on any site of the lattice, particles can
arbitrarily accumulate. One admits a single type $A$ of particles. 
Single particles can move diffusively, with rate $D$, to a nearest-neighbour site, irrespectively of its occupation. In addition,
particles are allowed to undergo the reactions \cite{Howa97,Paes04} 
\BEQ
A \to 2A \mbox{\rm ~~with rate $\mu_2$} \;\; , \;\; A \to \emptyset \mbox{\rm ~~with rate $\mu_0$}
\EEQ
and is called the {\em bosonic contact process with diffusion} ({\sc bcpd}). 
Clearly, for $\mu_2>\mu_0$, the whole lattice with be filled while for $\mu_2<\mu_0$ the particles will die out. 
There is a critical line $\mu_2=\mu_0$. The model leads to diffusion equations to average particle numbers and correlators. 
While the mean particle number $\langle \int\!\D\vec{r}\: a(t,\vec{r})\rangle=\rho_0$ 
stays constant on the critical line, one studies the connected correlator
\BEQ \label{gl:Cconn} 
C(t,s;\vec{r}) = \langle a(t\vec{r}_0)a(s,\vec{r}+\vec{r}_0)\rangle - \rho_0^2
\EEQ
The main issue is the comparison with non-equilibrium field-theory, since the action decomposes 
${\cal J}[\phi,\wit{\phi}]={\cal J}_0[\phi,\wit{\phi}]+{\cal J}_b[\phi,\wit{\phi}]$, where
\begin{subequations}
\begin{align}
{\cal J}_0[\phi,\wit{\phi}] &= \int \!\D t\int \!\D \vec{r} \: \wit{\phi} \bigl( 2{\cal M}\partial_t - \Delta_{\vec{r}} \bigr) \phi \label{gl:actionJ0} 
\\
{\cal J}_b[\phi,\wit{\phi}] &= -\mu_2 \int \!\D t \int \!\D\vec{r}\: \wit{\phi^2} \bigl( \phi + \rho \bigr) \label{gl:actionJb} 
\end{align}
\end{subequations}
These forms follow from probability conservation. 
The deterministic action ${\cal J}_0$ is the standard one, while the form of ${\cal J}_b$ 
is distinct from the ones considered so far. This will not modify the calculation of a 
response function $\langle \phi\wit{\phi}\rangle$. Because of the Bargman superselection rule, 
correlation functions must be found by expanding perturbatively the noisy part
${\cal J}_b$. This will give non-vanishing contribution of a correlator $\langle \phi\phi\rangle$ 
either from two applications of the first term in (\ref{gl:actionJb}) or else
a single application from the second term in (\ref{gl:actionJb}). 
It can be shown that the first contribution is less relevant than the second one \cite{Baum05b}. 
Then one may define an effective temperature $T_{\rm eff}=\mu_2\rho_0$ so that the leading terms are described by the same theory as exposed in 
section~\ref{sec:intro}. Then the scaling contributions are exactly those in (\ref{gl:ew}) \cite{Paes04,Baum05a}, 
up to normalisations and we know already that they agree with the predictions
(\ref{fonctions}), with $\nu=0$. The universal behaviour of the {\sc bcpd} is in the same universality class as the {\sc ew}-model, 
see table~\ref{tab:1}, which now includes the exact form of the 
scaling functions. However, corrections to scaling should be  distinct in both models.  

\subsection{Bosonic pair-contact process}

As our last example, we study the {\em bosonic pair-contact process with diffusion} {\sc bpcpd}. 
The model is defined in complete analogy with the previous example, but for a change in the
reactions \cite{Paes04} 
\BEQ
2A \to 3A \mbox{\rm ~~with rate $\mu_3$} \;\; , \;\; 2A \to \emptyset \mbox{\rm ~~with rate $\mu_0$}
\EEQ
Now the critical line is given by $2\mu_0=\mu_3$. Although the phases out-of-criticality are physically trivial 
(for $2\mu_0>\mu_3$ the particles die out and for $2\mu_0<\mu_3$ the whole lattice will be filled), 
the equations of motion only close, and hence be solved, at criticality 
and we shall restrict to this case from now on. 
As in the {\sc bcpd}, the mean particle number $\langle \int\!\D\vec{r}\: a(t,\vec{r})\rangle=\rho_0$ stays constant along the critical line. 
The behaviour along this line is determined by the parameter
\BEQ
\alpha = \frac{3\mu_3}{2D} ~~\;\; , \;\;~~ \frac{1}{\alpha_C} = 2 \int_0^{\infty} \!\D u\: \left( e^{-4u}I_0(4u) \right)^d 
\EEQ
and where the critical value $\alpha_C$ is finite for $d>2$. 
For $\alpha>\alpha_C$ the particles cluster on a single site \cite[Fig. 3.22]{Henk10} 
and the variance grows exponentially in time. For $\alpha\leq\alpha_C$, 
the spatial particle-distribution remains homogeneous. For $\alpha<\alpha_C$, 
the model is in the {\sc bcpd}-universality class. We discuss here the
multi-critical point at $\alpha=\alpha_C$. As in the {\sc bpcd}, we consider the connected correlator (\ref{gl:Cconn}). 
Mathematically, the key to the exact solution is the fact that 
the unconnected single-time correlator satisfies a self-consistency condition which is very close to the spherical constraints discussed above \cite{Paes04}. 
The single-time correlator was derived in \cite{Paes04}, 
two-time correlators (and responses) in \cite{Baum05a} whereas their interpretation in terms of local-scale-invariance was given in \cite{Baum05b}.  
As an example, we calculate the single-time correlator for $d>4$. Starting from \cite[(A.16)]{Paes04}, up to multiplicative constants and in the continuum limit
\BEA
C(t;{r}) &=& \int_0^{t} \!\D\tau' \int_0^{\tau'} \!\D\tau\; {\tau}^{-d/2} e^{-r^2/(4\tau)} 
              = \int_0^{t} \!\D\tau \int_{\tau}^t \!\D\tau'\;{\tau}^{-d/2} e^{-r^2/(4\tau)} 
\nonumber \\
&=& t^{-(d/2-2)}\, e^{-r^2/(4t)}\, U\left(2,\frac{d}{2};\frac{r^2}{4t}\right) 
\EEA
after exchanging the order of the integrations and with the identity \cite[(13.2.6)]{Abra65}.  

The interpretation, again via Janssen-de Dominicis non-equilibrium field theory, decomposes the action 
${\cal J}[\phi,\wit{\phi}]={\cal J}_0[\phi,\wit{\phi}]+{\cal J}_b[\phi,\wit{\phi}]$ into a deterministic part identical to (\ref{gl:actionJ0}) 
and a noise part which contains five distinct contributions. 
It has been shown that the most relevant one has the form given in (\ref{gl:actionJb}), where the rate now reads $\mu_3$  \cite{Baum05b}. 
Hence, we can re-apply the reasoning used for the {\sc bpcd} and can re-use the predictions (\ref{fonctions}) for the single-time and two-time observables.
The exactly derived results in the {\sc bpcpd} at the multi-critical point $\alpha=\alpha_C$ \cite{Paes04,Baum05a} 
fully agree with this expectation, with $\nu=0$. We include in
table~\ref{tab:1} the respective values for the non-equilibrium exponents, where the cases $2<d<4$ and $d>4$ need to be distinguished. 

Remarkably, the chemical-reaction systems of the {\sc bpcd} and {\sc bpcpd}, which do not satisfy detailed balance, 
reduce to the same kind of analysis as the spherical and voter models originally inspired from magnetism.

\section{Conclusions} \label{sec:conclusio}

\begin{table}[tb]  
\begin{center}
\begin{tabular}{|lrr|ccccc|}  \hline
\multicolumn{3}{|l|}{~} & \multicolumn{5}{r|}{~} \\[-0.3cm] 
\multicolumn{3}{|l|}{model}                      & ~$\delta=\wit{\delta}$~ & ~$\xi$~     & ~$\wit{\xi}$~ & ~$\wit{\delta}_2$~ & ~$\wit{\xi}_2$~ \\ \hline
\multicolumn{3}{|l|}{~} & \multicolumn{5}{r|}{~} \\[-0.3cm] 
Glauber-Ising         & ~$T=0$~~    & ~$d=1$~    & $1/4$                   & $0$         & $-1/2$        & $1/4$              & ~$-1/4$~     \\[0.2cm]
voter                 &             & ~$0<d<2$~  & $d/4$                   & $0$         & $d/2-1$       & $d/4$              & ~$(d-4)/4$~  \\
voter                 &             & ~$2<d$~    & $d/4$                   & $0$         & $0$           & $d/4$              & $0$          \\[0.2cm]
spherical             & ~$T=T_c$~   & ~$2<d<4$~  & $d/4$                   & ~$(4-d)/4$~ & ~$(d-4)/4$~   & $d/4$              & $(d-4)/4$    \\
spherical             & ~$T=T_c$~   & ~$4<d$~    & $d/4$                   & $0$         & $0$           & $d/4$              & $0$          \\[0.2cm]
{\sc ew} = {\sc bcpd} &             & ~$0<d$~    & $d/4$                   & $0$         & $0$           & $d/4$              & $0$          \\
{\sc bpcpd}           & ~$\alpha<\alpha_C$~  & ~$2<d$~   & $d/4$           & $0$         & $0$           & $d/4$              & $0$          \\[0.2cm]
Arcetri {\sc H}       & ~$T=T_c$~   & ~$0<d<2$~  & $d/4$                   & $(2-d)/4$   & $(d-2)/4$     & $d/4$              & $(d-2)/4$    \\
Arcetri {\sc H}       & ~$T=T_c$~   & ~$2<d$~    & $d/4$                   & $0$         & $0$           & $d/4$              & $0$          \\[0.2cm] 
{\sc bpcpd}           & ~$\alpha=\alpha_C$ & ~$2<d<4$~   &  $d/4$          & $0$         & $0$           & $d/4$              & $(d-2)/4$    \\ 
{\sc bpcpd}           & ~$\alpha=\alpha_C$ & ~$4<d$~     &  $d/4$          & $0$         & $0$           & $d/4$              & $1/2$        \\ \hline 
\end{tabular}\end{center}
\caption[tab1]{Values of the scaling dimensions needed in the calculation of correlators and responses 
in exactly solved models of non-equilibrium critical dynamics with $\mathpzc{z}=2$. 
{\sc ew} denotes the Edwards-Wilkinson model \cite{Edwa82}, {\sc bcpd} and {\sc bpcpd} 
are the diffusive contact and pair contact processes \cite{Paes04}, considered on their critical line. 
\label{tab:1}
}
\end{table}

The new non-equilibrium and time-dependent representation of the Schr\"odinger algebra \cite{Henk25,Henk25c} 
has been used to predict the exact form of single-time and two-time correlators 
(we include more early results on the two-time responses) for systems undergoing non-equilibrium dynamics 
after a quench onto criticality, from a fully disordered initial state. 
These predictions (\ref{fonctions}) for the entire scaling functions only depend on a small number of non-equilibrium exponents, 
notably $\wit{\delta}_2$ and $\wit{\xi}_2$. These characterise the composite field $\wit{\phi^2}$, which describes the coupling of the system to its environment. 
Their values in various models are collected in table~\ref{tab:1}. In all cases, we have found $\nu=0$. 
Comparison of the results of the voter, spherical and Arcetri models shows that the {\sc ew} 
universality class acts as the mean-field theory for all of them, although that
mean-field regime is reached for different values of the dimension $d$ in the various models. 
The multi-critical {\sc bpcpd} at $\alpha=\alpha_C$ has its own, distinct, mean-field theory. 
The non-trivial nature of these models for $d<d^*$ is not always apparent in the values of exponents such as 
$\lambda$ or $\mathpzc{z}=2$, but rather in the form of the
(connected) correlator scaling functions. The existing large variety of models 
suggests that further exactly solved systems \cite{Touz23} within this framework might exist. 
Possible extensions to quantum (hydro-)dynamics \cite{Zahr25} would be of great interest. 

\noindent
{\bf Acknowledgements:} This work was supported by the french ANR-PRME UNIOPEN (ANR-22-CE30-0004-01), 
PHC RILA (51305UC/KP06-Rila/7) and Bulgarian National Science Fund, grant KP-06-N88/3.

\end{document}